\documentclass[graybox,natbib,nosecnum]{svmult}
\bibpunct{(}{)}{;}{a}{}{,} 

\pdfoutput=1   

\usepackage{mathptmx}       
\usepackage{helvet}         
\usepackage{courier}        
\usepackage{type1cm}        

\usepackage{makeidx}         
\usepackage{graphicx}        
\usepackage{multicol}        
\usepackage[bottom]{footmisc}
\usepackage[normalem]{ulem}	
\usepackage{hyperref}  

\usepackage{soul}   
\usepackage{tikz}
\newcommand*\aap{A\&A}

\newcommand*\apj{ApJ}
\newcommand*\apjl{ApJ}

\newcommand*\icarus{Icarus}

\newcommand*\mnras{MNRAS}

\makeindex             

\begin{document}

\title*{Trojan Exoplanets}
\author{Philippe Robutel  and Adrien Leleu }
\institute{Philippe Robutel \at IMCCE, Observatoire de Paris, PSL University, CNRS, Sorbonne
Universit\'e, 77 avenue Denfert-Rochereau, 75014 Paris, France, \email{philippe.robutel@obspm.fr}
\and Adrien Leleu \at Observatoire de Gen\`eve, Universit\'e de Gen\`eve, Chemin Pegasi, 51, 1290 Versoix, Switzerland, \email{adrien.leleu@unige.ch}}
%
%
\maketitle

{\it To be published in: Handbook of Exoplanets, 2nd Edition, Hans Deeg and Juan Antonio Belmonte (Eds. in Chief), Springer International Publishing AG, part of Springer Nature.
}
\vspace{2 cm}

\abstract{Co-orbital exoplanets are a by-product of the models of formation of planetary systems. However, none have been detected in nature thus far. Although challenging, the observation of co-orbital exoplanets would provide valuable information on the formation of planetary systems as well as on the interactions between planets and their host star.
After a brief review of the stability and formation issues of co-orbital systems, some observational methods dedicated to their detection are presented.
}

\section{Introduction}

The existence of Trojans in extrasolar systems has already been discussed in the chapter “Special Cases: Moons, Rings, Comets, and Trojans”. The present chapter generalizes this concept by considering {co-orbital} planets.  This kind of configuration consists of a planet pair trapped in the 1:1 {mean motion resonance}.   Thus, contrary to the Trojan case, discussed in the above-mentioned chapter, for which the mass is supposed to be small enough not to perturb the motion of the two other bodies (restricted case), it is possible to have co-orbital planets of comparable masses provided their sum remains small with respect to the mass of the star (less than $25$ times smaller). 

Co-orbital configurations are well known in the solar system. More than a century after the pioneer works of \citep{Euler1764}, who discovered the aligned configurations in the general 3-body problem, and  \cite{Lagrange1772} who demonstrated the existence of the famous equilateral configurations, was discovered Achilles, the first Trojan asteroid \citep{Wolf1906}.  Since then, thousands of small bodies are observed in the Jovian {Trojan} swarms $L_4$ and $L_5$ as well as some Neptune, Uranus, Mars and Earth Trojans (see \url{https://www.minorplanetcenter.net/iau/lists/Trojans.html}). Saturn's satellite system also hosts co-orbital satellites, in particular the famous Janus and Epimetheus pair which, unlike most Trojans in the solar system, follow horseshoe trajectories of very large amplitude \citep{DeMu1981b,ThomasBurnsHedmanHelfensteinMorrison2013}.

Although many theoretical works suggest that co-orbital exoplanets may also exist, such a pair of planets has yet to be observed. In particular, out of the hundreds of multi-planetary systems discovered by the Kepler mission, none were found in co-orbital configuration thus far. Although this result points to a relative rarity of the configuration, part of this absence might be due to observational biases.

After reviewing the stability issues of co-orbital systems, the main methods dedicated to their detection  will be outlined.

\section{Co-orbital configurations and their stability}
\label{sec:stabi}

\begin{figure}
\includegraphics[height=300pt]{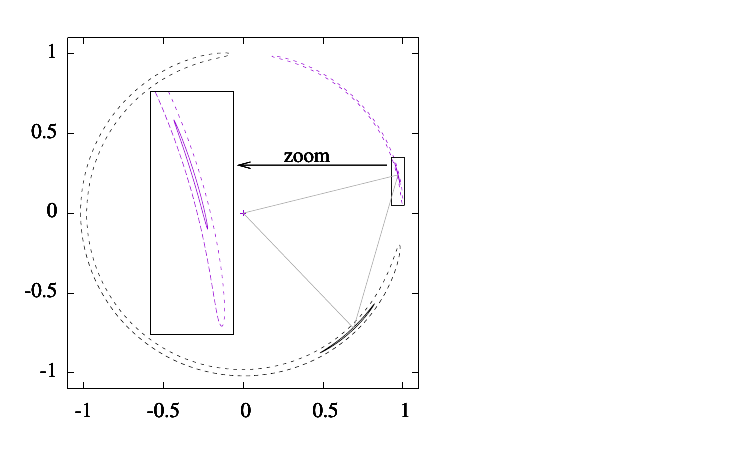}
\caption{Tadpole (solid lines) and horseshoe orbits (dashed lines) seen in a rotating frame with the common orbital frequency $n$ of both planets. The coordinates of the central body are $(0,0)$ while while the co-orbital's orbits are plotted in $\left(a_j\cos(\lambda_j-n\,t),a_j\sin(\lambda_j-n\,t)\right)$. The purple orbits correspond to the planet $1$ whose mass satisfies $m_1/m_0 = 10^{-4}$ while those of the second planet with $m_2/m_0 = 3\, 10^{-5}$ are plotted in purple. The grey equilateral triangle indicates the Lagrange configuration.
The initial conditions of the horseshoe orbit satisfy $a_1=a_2=1$, $e_1=e_2=0$, $\lambda_1 - \lambda_2 = \varpi_1-\varpi_2 =  15^\circ$, while $\lambda_1 - \lambda_2 = \varpi_1-\varpi_2 =  45^\circ$ is set for tadpole.}
\label{fig:HS}
\end{figure}

The first stability result was established by  \cite{Ga1843} who proved that the Lagrange equilateral configuration was stable, in the coplanar circular case,  providing the masses $m_j$ satisfy the following relation 
$$
(m_0+m_1+m_2)^2 \geq 27(m_0m_1 +  m_0m_2 + m_1m_2).
$$
   In order to be satisfied, this condition requires that the mass of one body, for instance $m_0$, dominates the others. Thus, it can be verified that, the mass ratio $\mu = (m_1 +m_2)/m_0$ must remain smaller than $0.04$ to ensure stability. Therefore, it is not necessary for one of the masses to be negligible in front of the other two, as is the case with the Jovian Trojans.  In particular, there is no reason why a Lagrange configuration consisting of one solar mass star and two Jovian planets should not be stable. 
These particular periodic solutions being linearly stable, their neighborhoods are filled with quasi-periodic orbits, known as tadpole-orbits, or also trojan-orbits by analogy with the Jovian Trojans , that correspond to small to moderate deformations of the equilateral configurations (Fig. \ref{fig:HS}, solid lines). Although the Lagrangian triangular equilibria are linearly stable, the overall stability of the tadpole region is not guaranteed.  Indeed, for some particular values of the mass ratio $\mu$ associated with resonances, the stability domain shrinks, or even disappears entirely. \citep[e.g.][]{Dan1964,GiDeFoGaSi1989,Robe2002,Nauenberg2002,ErNaSaSuFro2007}.  For a given tadpole orbit, the libration angle $\zeta$, equal to the difference of the planetary mean-longitude $\lambda_1$ and $\lambda_2$, oscillates around a value close to $\pm 60^\circ$ (depending on whether one considers the $L_4$ or $L_5$ neighborhood)  with a frequency $\nu$ equal to $n\sqrt{27\mu}/2$ very close to Lagrange equilibria, $n$ being the orbital frequency common to the two planets, and remains of the same order throughout the tadpole areas. In case of significant eccentricities, perihelia slowly precess with small secular frequency of order $\mu$, which is therefore small compared to the libration frequency.

\begin{figure}
\includegraphics[height=280pt]{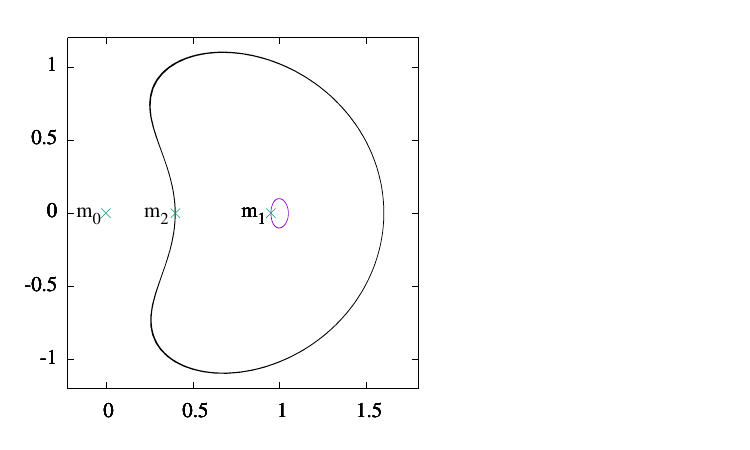}

\caption{Quasi-satellite orbit. Trajectory of the first planet (purple) and of the second one (black)  seen in a rotating frame with the common orbital frequency $n$ of the both planets.  The masses are the same than in Fig. \ref{fig:HS}, while the initial elliptic elements satisfy $a_1=a_2=1$, $e_1=0.05, e_2 = 0.6$. }
\end{figure}

When $\mu$  is lower than a specific value close to $0.0004$ \citep{LauCha2002}, the libration angle $\zeta$ can undergo variations of much greater amplitude given birth  to horseshoe orbits, which become stable below this critical value (Fig. \ref{fig:HS}, dashed lines).
Along horseshoe orbits, the libration angle $\zeta$ evolves with an amplitude which is greater that  $312^\circ$, while for tadpole orbits this amplitude is bounded by $156^\circ$.    
As with the tadpole orbits, the coplanar dynamic is driven by three distinct time scales: the fast one, related to the orbital period; the semi-fast, associated with the libration inside the 1:1  mean-motion resonance, and the secular related to orbital precession. 
For large eccentricities, the width of the tadpole and horseshoe domains shrink to make room for quasi-satellites orbits \citep{GiuBeMiFe2010,PoRoVi2017}. Other families of orbits also exist for highly eccentric orbits \citep{LeRoCo2017}. 
As it is the case in the solar system, it can also be assumed that the orbits of the two co-orbital are not coplanar.
If the mutual inclination is low, the main dynamical features are very similar to those described above. Higher inclinations allow transitions between theses different types of orbits and merging between them \citep{Namouni1999}, but in any case, the mutual inclination of stable co-orbital tadpole-orbits may not exceed about $60^\circ$ degrees \citep[see][]{QiRu2020}.
Other kinds of trajectories, which are not discussed here, also exist within the co-orbital resonance such as retrograde motions \citep{MoNa2016,SiArNeZe2013}.

When the co-orbital bodies are embedded in a multi-planetary system, the dynamics become much more complex and the size of the stability regions may decrease significantly. Because of the additional fundamental frequencies associated with the whole planetary system (three frequencies per planet) a wide range of resonances can destabilize the coorbital system \citep{MaTriSch2003,RoGa2006,Leleu2019}. These phenomena have many consequences. They are the cause of the slow erosion of the Jovian Trojan swarms \citep{LevisonSS97,RoGa2006}, they introduce unstable areas in the tadpole and horseshoe regions \citep{Leleu2019}, and can go so far as to make the whole coorbital region unstable, as is the case with Saturn \citep{HoWi1993,NeDo02,HoScLi2014}.

\section{Formation and evolution of co-orbital exoplanets}
\label{sec:evo}

To understand the particularities of the co-orbital resonance, it is useful to recall the current understanding for the formation and evolution of other resonant configurations.  

First and second-order MMRs are natural and common outcomes of models of formation of planetary systems: as planets form in the protoplanetary disc, they tend to migrate toward the star. Depending on their relative migration speed and eccentricities, they have a probability of being captured in each of the MMRs they cross \citep[e.g.][]{LePe2002}, the most commonly observed being the 3:2 MMR. Models hence predict that toward the end of the protoplanetary disc phase, numerous planetary systems are in close-in, compact resonant state \citep[e.g.][]{CreNe2008,Coleman2019,NGPPS1}. As the protoplanetary disc dissipates, the eccentricity damping lessens, which can lead to instabilities \citep[e.g.][]{TePa2007,PuWu2015,IzOgRaMo2017}. For planets that remained in resonance, and are close enough to their host star, tides become the dominant force that affect the architecture of the systems, which can then lead to a departure of the period ratio from the exact resonance \citep{HenLe1983,PaTe2010,DeLaCoBo2012}. This succession of dissipative processes partly explain why the observed close-in multi-planetary systems are often just outside MMRs, but rarely inside \citep{Fabrycky2014}.

Both of the aforementioned phenomenon, i.e. the smooth convergent migration in resonance during the protoplanetary disc, and the slow departure from resonance due to tides, cannot be applied to the co-orbital resonance. Indeed, this resonance is surrounded by a chaotic area due to the overlap of first-order MMRs \citep{Wi1980,DePaHo2013,PeLaBo2018}, and a slow crossing of this area would generally result in the excitation of the bodies' eccentricities, leading to collision or scattering instead of the capture into the resonance. 

Nevertheless, several formation scenarios of co-orbital configurations can be found in the literature. \cite{LauCha2002} suggested planet-planet gravitational scattering followed by strong eccentricity damping due to the protoplanetary disc, or accretion \textit{in situ} at the $L_4/L_5$ points of a primary. 
Alternatively, the presence of an external body such as a massive planet can help the capture of co-orbital bodies through resonant crossing \citep[see][for the case of the solar system ]{NeVoMo2013}.

In the \textit{in situ} scenario, different studies yielded different upper limit to the mass that can form at the $L_4/L_5$ equilibrium point of a giant planet: \citet{BeSa2007} obtained a maximum mass of $\sim 0.6 M_\oplus$, while \citet{LyJo2009} obtained $5 -15 M_\oplus$ planets in the tadpole area of a Jupiter-mass primary.

Co-orbital pairs form naturally in models of formation of planetary systems \citep[e.g.][]{CreNe2009,Coleman2019,NGPPS1} in typically a few percent of the systems, through scattering, but also during the migration and encounter of resonant chains: in the simulation of the formation of Trappist-1 by \cite{Coleman2019}, co-orbitals were mostly found as a part of a resonant chain, which appears to have a stabilising effect \citep{LeCoAt2019}.
However, models of the formation of planetary systems model the disc-planet interactions using analytical expression computed for single planets. The presence of a co-orbital configuration has been shown to impact significantly the local proto-planetary disc structure, sometime changing the stability of the co-orbital pair \citep{PiRa2014,LeCoAt2019}. 

Nonetheless, co-orbitals can also form in hydro-dynamical simulations of the proto-planetary disc with two or more planets embedded \citep[e.g.][]{Crida2009,PeAtKl2019}. In several simulations of the formation of the outer part of the solar system, \cite{Crida2009} found that Uranus and Neptune ended up in co-orbital configuration, while being both trapped in the same MMR with Saturn. This reinforces the idea that resonant chains might be favorable to the formation of co-orbital configuration. As long as the planets do not fully open a gap in the proto-planetary disc (typically for planet smaller than Neptune), co-orbitals evolving in the proto-planetary disc appear to be more stable when the leading planet is more massive \citep{PiRa2014,LeCoAt2019}, while trailing more massive planet tends to disrupt the configuration. In the case of a terrestrial planet co-orbital with a Jupiter-sized planet, the stability seemed to mainly depend on the proto planetary disc parameters.

For close-in systems, the effect of tides raised by the star onto co-orbital planets appears to always be disruptive \citep{RoGiMi2013,CouRoCo2021,DoLi2022}. However this phenomenon only impacts planets close to their host star, and most co-orbitals that are at more than 10 days of orbital period at the disc dispersal should survive for billions of years \citep{CouRoCo2021}.

\section{Detection of co-orbital planets}

Stacking the phase-folded lightcurve of thousands of KOIs, \cite{HiAn2015} found statistical hints of dips near the $L_4$ and $L_5$ equilibria of known planets. However, despite the discovery of thousands of exoplanets in hundreds of multiplanetary systems, no extra-solar co-orbital configurations has been confirmed thus far. The subsequent sections discuss the signature of co-orbital configurations in photometric (transit) and spectroscopic (radial velocity) measurements.

\subsection{Transit}
\label{sec:TTVbias}

Transit surveys such as {Kepler}, {TESS }and the upcoming {PLATO} mission have discovered the bulk of known exo-planetary systems. As for any other pair of planets, the detection of co-orbitals by transit would require both planets to orbit roughly in the same orbital plane. In addition, their detection requires that each of the co-orbital is large enough for its individual transits to be detected. 

For smaller planets, the data-reduction pipelines of transit surveys use an approach similar to the box least square algorithm \citep[BLS, see][]{Kovacs2002,Jenkins2010,Jenkins2016}. These algorithms fold each light curve over a large number of different periods and look for transits in the folded data.

\begin{figure}[!ht]
\begin{center}
\includegraphics[width=0.99\textwidth]{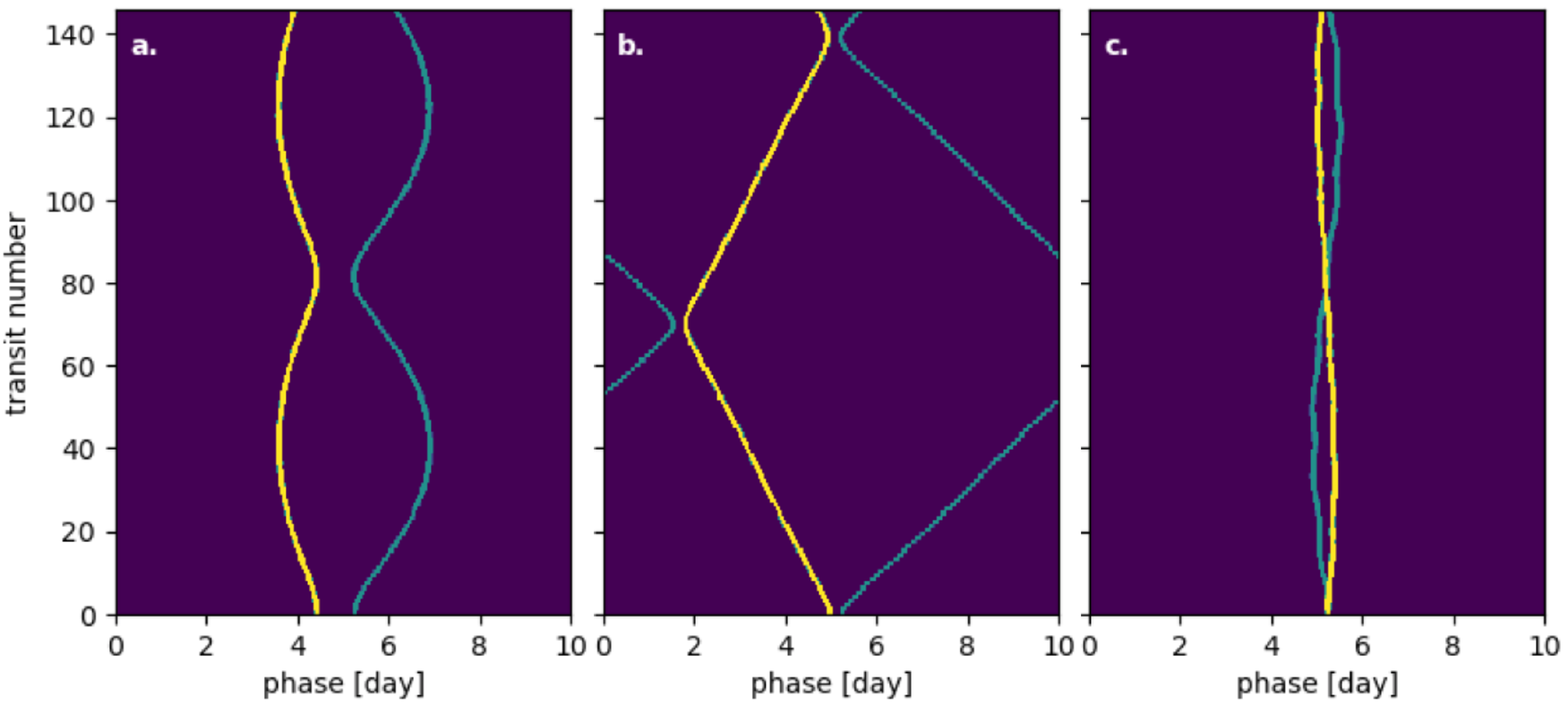}
\caption{\label{fig:coorb_TTV} River diagrams \citep{Carter2012} for co-orbitals in tadpole orbit (a.), horseshoe orbit (b.) and quasi-satellite (c.). These diagrams show the transit timings of each planet as function of the phase of the average period of the planets (here 10 days, x axis), for each epoch (y axis). In each case, the yellow track shows the transit of the most massive of the two planets, and the blue track shows the transits of the least massive. Assuming a sun-like star, $m_1=2m_2=10m_\oplus$ for a. and c.; and $m_1=2m_2=2m_\oplus$ for b.}
\label{fig:rivers}
\end{center}
\end{figure}

This method is adapted to planets on unperturbed or moderately perturbed orbits for which the transits occur periodically. However, gravitational interactions between planets in resonant systems produce {transit timing variations} (TTVs). Figure \ref{fig:rivers} shows examples of TTVs for different co-orbital configurations. These TTVs can prevent the detection of small planets if their amplitude $\sigma_{TTV}$ (amplitude of the oscillation of the curves along the x-axis) is larger than the transit duration $D$ of the planet (width of the curves), and their period is shorter than the mission duration (y-axis). In this case, the signal-to-noise ratio of the transit of the planet recovered by BLS-like algorithms decreases by a factor $ 1/(1+\sigma_{TTV}/D)$ \citep{RIVERS1}. In addition, the recovered transit shape is altered by the TTVs, and therefore the planet can be missed, or mistaken for a false positive \citep{RIVERS2}.

Assuming that the masses $m_p$ of the two planets are comparable and that it is the same for their eccentricities $e_p$, a rough calculation \citep[see][and references therein for accurate expressions]{GaBeGi2021,NeVo2016} shows that, for a MMR of order $q$, the TTVs occur on a timescale proportional to:
\begin{equation}
P_{TTV} \propto (m_p / m_\star)^{-1/2} e_p^{-q/2} P_{orb} \, .
\label{eq:TTVres}
\end{equation}
Note that this formula is not singular because the eccentricities cannot be cancelled simultaneously if $q \geq 1$. 
With $q=0$, co-orbitals hence generate faster TTVs than other MMRs. In addition their amplitude  can be a significant fraction of the orbital period of the planets \citep[e.g.][ and Fig. \ref{fig:coorb_TTV}]{NeVo2014,Leleu2019}. Co-orbitals are hence especially subject to the TTV bias, and dedicated methods are required to recover potential pairs of small co-orbitals in transit surveys data.


Assuming a large mass discrepancy between co-orbitals, \cite{janson2013} looked for dips near the $L_4$ and $L_5$ equilibrium of known planets, using an analytical model for small libration of the resonant angle and low eccentricities. No significant signal was found, despite a sensitivity level down to Earth-size objects. However, such a signal could still exist for larger amplitudes of libration or eccentricities. 

If the co-orbitals are similar in size, both of them could be affected by the TTV bias, in which case both would have been missed by BLS-like approaches, and all orbital periods have to be probed to recover them. The QATS algorithm is adapted to the research of planets with TTVs: it allows the timing of each transit to vary at the same time that it adjusts the best transit depth and duration \citep{CarAg2013}. Alternatively, neural networks trained to recover the track of planets in river diagrams were shown to be capable of detecting planets with large TTVs and low S/N \citep{RIVERS1,RIVERS2}.

Co-orbitals might also be detected even if only one of them transit: with a good phase coverage and high S/N transits, it is possible, based on the TTVs of a single planet, to determine in which MMR that planet and its perturber are \citep[see the case of KOI-142,][]{Nesvorny2013}. When co-orbitals librate with a significant amplitude the peculiar shape of their TTVs \citep[arches for tadpole-orbits as shows in Fig. \ref{fig:rivers}-a and  triangle for horseshoe as  shown in Fig. \ref{fig:rivers}-b and in the figure 8 by][]{VoNe2014} could suffice to confirm the existence of a co-orbital configuration.

\subsection{Radial velocity}

Co-orbitals librating in Trojan or horseshoe configurations modulate the radial velocity (RV) signal with a period equal to their libration period. The amplitude of this modulation increases when the mass of the planets tend toward equal value and the amplitude of libration increases \citep{LauCha2002,FoHo2007}. If this modulation is observed, the estimation of the Keplerian signal and the first modulation harmonics is sufficient to recover the orbit and masses of the planets \citep{LeRoCo2015}.

For a pair of co-orbital at the exact 
 Lagrangian equilibrium, the modulation vanishes and the generated radial velocities are identical from those induced by a single planet located at their longitudinal barycenter with a mass of 
$\sqrt{m_1^2 + m_2^2 + m_1m_2} , $
%
 \citep[see][]{GiuBe2012,LeRoCoLi2017}, and hence the co-orbital configuration cannot be identified.
 However, in the best configuration (equal mass planets, large libration amplitude), the maximum modulation amplitude can be up to $\sim 30\%$ for tadpole orbit and $100\%$ for horseshoe (the RV signal vanishes when the planets are in opposition with respect to the star). In this case, the modulation would be detectable with current radial-velocity instruments. It however requires a long observational baseline to cover the libration period, and might be difficult to disentangle from other dynamical effects or noise. In the case of quasi-satellite orbits, \cite{LauCha2002} remark that the RV signal is also similar to one induced by a single planet, however a small fraction of the orbital phase shows a strong variation in RV.

The co-orbital signature is similar in astrometry. However, the system observed in astrometry tend to be at larger orbital period, making it unlikely to observe a system long enough to see the resonant modulation of the signal.

\subsection{Radial velocity and transit}

Combining RVs and transits allows observers to detect co-orbitals when only one of them transits and without the need to observe the resonant libration. 
For a single planet in circular orbit, the time of mid-transit coincides with the instant where the radial-velocity reaches its mean value. If the transiting planet has a co-orbital companion, there is a time shift $\Delta t$ between 
these two phenomena that depends on the mass distribution between the co-orbitals and their angular separation $\zeta$ \citep{FoGa2006,LeRoCoLi2017}.
If $P$ is the common orbital period, this time shift reads:
%
\begin{equation}
\Delta t = \frac{P}{2\pi}\left( \frac{\zeta}{2} - \arctan\left( 
(1 - 2\delta)\tan\frac{\zeta}{2} 
\right) \right) \, {\rm with } \quad \delta = \frac{m_2}{m_1+m_2} ,
\end{equation}

$m_1$ being the transiting planet mass and $m_2$ the co-orbital one. 
But this time shift also exits with a single planet on an elliptic orbit of eccentricity $e$ and argument of pericentre $\omega$. In this case, the time shift reads:
\begin{equation}
\Delta t = -e\cos\omega \frac{P}{2\pi} \, .
\end{equation}
The sensibility of this approach hence depends on the precision reached on the quantity $e\cos\omega$ of the transiting planet. So far, this approach could only be used on hot Jupiter, for which the observation of the secondary eclipse can constrain $e\cos\omega$ \citep{MaWi2009,LiBo2017,LiBo2018}. No co-orbital companion were found in these studies, which is consistent with their instability to tides \citep{CouRoCo2021}.

\section{Conclusion}

Co-orbital exoplanets can be stable for duration longer than the life of their host star if they are sufficiently far away from the star to prevent large tidal disruption, and in the absence of large perturbations by additional planets. In addition, models of formation of planetary systems tend to create pairs of co-orbital exoplanets. 
However this resonance has several particularities, like that of being surrounded by  large chaotic zones, or to modify the local proto-planetary disc structure because of the presence of two planets at the same distance to the star. This require extra care when studying to formation and evolution of co-orbital systems.

Co-orbital exoplanets have not be found yet. However, the methods used to detect the bulk of the exoplanets  have biases that impede their capability to recover exoplanets in this kind of configuration. The development of dedicated data analysis methods is required, and the continuous increase in measurement precision and observation baseline could allow us to identify a co-orbital signature in the future. Co-orbitals could also be found by other methods such as direct imaging which, combined with stability studies, has been shown to be able to identify the resonant state of multi-planetary systems \citep{Gozdziewski2020}.   

%
%

\bibliography{biblio.bib}

\begin{thebibliography}{}

\bibitem[{Beaug{\'e}} et~al., 2007]{BeSa2007}
{Beaug{\'e}}, C., {S{\'a}ndor}, Z., {{\'E}rdi}, B., and {S{\"u}li}, {\'A}.
  (2007).
\newblock {Co-orbital terrestrial planets in exoplanetary systems: a formation
  scenario}.
\newblock {\em \aap}, 463:359--367.

\bibitem[{Carter} and {Agol}, 2013]{CarAg2013}
{Carter}, J.~A. and {Agol}, E. (2013).
\newblock {The Quasiperiodic Automated Transit Search Algorithm}.
\newblock {\em \apj}, 765(2):132.

\bibitem[{Carter} et~al., 2012]{Carter2012}
{Carter}, J.~A., {Agol}, E., {Chaplin}, W.~J., {Basu}, S., {Bedding}, T.~R.,
  {Buchhave}, L.~A., {Christensen-Dalsgaard}, J., {Deck}, K.~M., {Elsworth},
  Y., {Fabrycky}, D.~C., {Ford}, E.~B., {Fortney}, J.~J., {Hale}, S.~J.,
  {Handberg}, R., {Hekker}, S., {Holman}, M.~J., {Huber}, D., {Karoff}, C.,
  {Kawaler}, S.~D., {Kjeldsen}, H., {Lissauer}, J.~J., {Lopez}, E.~D., {Lund},
  M.~N., {Lundkvist}, M., {Metcalfe}, T.~S., {Miglio}, A., {Rogers}, L.~A.,
  {Stello}, D., {Borucki}, W.~J., {Bryson}, S., {Christiansen}, J.~L.,
  {Cochran}, W.~D., {Geary}, J.~C., {Gilliland}, R.~L., {Haas}, M.~R., {Hall},
  J., {Howard}, A.~W., {Jenkins}, J.~M., {Klaus}, T., {Koch}, D.~G., {Latham},
  D.~W., {MacQueen}, P.~J., {Sasselov}, D., {Steffen}, J.~H., {Twicken}, J.~D.,
  and {Winn}, J.~N. (2012).
\newblock {Kepler-36: A Pair of Planets with Neighboring Orbits and Dissimilar
  Densities}.
\newblock {\em Science}, 337:556.

\bibitem[{Coleman} et~al., 2019]{Coleman2019}
{Coleman}, G.~A.~L., {Leleu}, A., {Alibert}, Y., and {Benz}, W. (2019).
\newblock {Pebbles versus planetesimals: the case of Trappist-1}.
\newblock {\em \aap}, 631:A7.

\bibitem[{Couturier} et~al., 2021]{CouRoCo2021}
{Couturier}, J., {Robutel}, P., and {Correia}, A.~C.~M. (2021).
\newblock An analytical model for tidal evolution in co-orbital systems. i.
  application to exoplanets.
\newblock {\em Celest. Mech. Dyn. Astron.}

\bibitem[{Cresswell} and {Nelson}, 2008]{CreNe2008}
{Cresswell}, P. and {Nelson}, R.~P. (2008).
\newblock {Three-dimensional simulations of multiple protoplanets embedded in a
  protostellar disc}.
\newblock {\em aap}, 482:677--690.

\bibitem[{Cresswell} and {Nelson}, 2009]{CreNe2009}
{Cresswell}, P. and {Nelson}, R.~P. (2009).
\newblock {On the growth and stability of Trojan planets}.
\newblock {\em Astron. Astrophys.}, 493:1141--1147.

\bibitem[{Crida}, 2009]{Crida2009}
{Crida}, A. (2009).
\newblock {Minimum Mass Solar Nebulae and Planetary Migration}.
\newblock {\em \apj}, 698(1):606--614.

\bibitem[{Danby}, 1964]{Dan1964}
{Danby}, J.~M.~A. (1964).
\newblock Stability of the triangular points in the elliptic restricted problem
  of three bodies.
\newblock {\em Astron. Astrophys.}, 69:165.

\bibitem[{Deck} et~al., 2013]{DePaHo2013}
{Deck}, K.~M., {Payne}, M., and {Holman}, M.~J. (2013).
\newblock {First-order Resonance Overlap and the Stability of Close Two-planet
  Systems}.
\newblock {\em \apj}, 774:129.

\bibitem[{Delisle} et~al., 2012]{DeLaCoBo2012}
{Delisle}, J.-B., {Laskar}, J., {Correia}, A.~C.~M., and {Bou{\'e}}, G. (2012).
\newblock {Dissipation in planar resonant planetary systems}.
\newblock {\em Astron. Astrophys.}, 546:A71.

\bibitem[{Dermott} and {Murray}, 1981]{DeMu1981b}
{Dermott}, S.~F. and {Murray}, C.~D. (1981).
\newblock {The dynamics of tadpole and horseshoe orbits II. The coorbital
  satellites of saturn}.
\newblock {\em Icarus}, 48:12--22.

\bibitem[{Dobrovolskis} and {Lissauer}, 2022]{DoLi2022}
{Dobrovolskis}, A.~R. and {Lissauer}, J.~J. (2022).
\newblock {Do tides destabilize Trojan exoplanets?}
\newblock {\em \icarus}, 385:115087.

\bibitem[{Emsenhuber} et~al., 2021]{NGPPS1}
{Emsenhuber}, A., {Mordasini}, C., {Burn}, R., {Alibert}, Y., {Benz}, W., and
  {Asphaug}, E. (2021).
\newblock {The New Generation Planetary Population Synthesis (NGPPS). I. Bern
  global model of planet formation and evolution, model tests, and emerging
  planetary systems}.
\newblock {\em \aap}, 656:A69.

\bibitem[{{\'E}rdi} et~al., 2007]{ErNaSaSuFro2007}
{{\'E}rdi}, B., {Nagy}, I., {S{\'a}ndor}, Z., {S{\"u}li}, {\'A}., and
  {Fr{\"o}hlich}, G. (2007).
\newblock {Secondary resonances of co-orbital motions}.
\newblock {\em MNRAS}, 381:33--40.

\bibitem[Euler, 1764]{Euler1764}
Euler, L. (1764).
\newblock Considerationes de motu corporum coelestium.
\newblock {\em Novi commentarii academiae scientiarum Petropolitanae. Berlin
  acad.}, 10:544--558.

\bibitem[{Fabrycky} et~al., 2014]{Fabrycky2014}
{Fabrycky}, D.~C., {Lissauer}, J.~J., {Ragozzine}, D., {Rowe}, J.~F.,
  {Steffen}, J.~H., {Agol}, E., {Barclay}, T., {Batalha}, N., {Borucki}, W.,
  {Ciardi}, D.~R., {Ford}, E.~B., {Gautier}, T.~N., {Geary}, J.~C., {Holman},
  M.~J., {Jenkins}, J.~M., {Li}, J., {Morehead}, R.~C., {Morris}, R.~L.,
  {Shporer}, A., {Smith}, J.~C., {Still}, M., and {Van Cleve}, J. (2014).
\newblock {Architecture of Kepler's Multi-transiting Systems. II. New
  Investigations with Twice as Many Candidates}.
\newblock {\em apj}, 790:146.

\bibitem[{Ford} and {Gaudi}, 2006]{FoGa2006}
{Ford}, E.~B. and {Gaudi}, B.~S. (2006).
\newblock {Observational Constraints on Trojans of Transiting Extrasolar
  Planets}.
\newblock {\em apjl}, 652:L137--L140.

\bibitem[{Ford} and {Holman}, 2007]{FoHo2007}
{Ford}, E.~B. and {Holman}, M.~J. (2007).
\newblock {Using Transit Timing Observations to Search for Trojans of
  Transiting Extrasolar Planets}.
\newblock {\em apjl}, 664:L51--L54.

\bibitem[{Gallardo} et~al., 2021]{GaBeGi2021}
{Gallardo}, T., {Beaug{\'e}}, C., and {Giuppone}, C.~A. (2021).
\newblock {Semianalytical model for planetary resonances. Application to
  planets around single and binary stars}.
\newblock {\em \aap}, 646:A148.

\bibitem[{Gascheau}, 1843]{Ga1843}
{Gascheau}, G. (1843).
\newblock Examen d'une classe d'{\'e}quations diff{\'e}rentielles et
  application {\`a} un cas particulier du probl{\`e}me des trois corps.
\newblock {\em C. R. Acad. Sci. Paris}, 16(7):393--394.

\bibitem[Giorgilli et~al., 1989]{GiDeFoGaSi1989}
Giorgilli, A., Delshams, A., Fontich, E., Galgani, L., and Sim{\'o}, C. (1989).
\newblock Effective stability for a hamiltonian system near an elliptic
  equilibrium point, with an application to the restricted three body problem.
\newblock {\em JDIFE}, 77(1):167--198.

\bibitem[{Giuppone} et~al., 2010]{GiuBeMiFe2010}
{Giuppone}, C.~A., {Beaug{\'e}}, C., {Michtchenko}, T.~A., and {Ferraz-Mello},
  S. (2010).
\newblock Dynamics of two planets in co-orbital motion.
\newblock {\em MNRAS}, 407:390--398.

\bibitem[{Giuppone} et~al., 2012]{GiuBe2012}
{Giuppone}, C. A.~{Benitez-Llambay}, P., , and {Beaug{\'e}}, C. (2012).
\newblock Origin and detectability of co-orbital planets from radial velocity
  data.
\newblock {\em MNRAS}.

\bibitem[Goździewski and Migaszewski, 2020]{Gozdziewski2020}
Goździewski, K. and Migaszewski, C. (2020).
\newblock An exact, generalized laplace resonance in the hr 8799 planetary
  system.
\newblock {\em The Astrophysical Journal}, 902(2):L40.

\bibitem[{Henrard} and {Lemaitre}, 1983]{HenLe1983}
{Henrard}, J. and {Lemaitre}, A. (1983).
\newblock {A second fundamental model for resonance}.
\newblock {\em Celestial Mechanics}, 30:197--218.

\bibitem[{Hippke} and {Angerhausen}, 2015]{HiAn2015}
{Hippke}, M. and {Angerhausen}, D. (2015).
\newblock {Photometrys Bright Future: Detecting Solar System Analogs with
  Future Space Telescopes}.
\newblock {\em The Astrophysical Journal}, 810:29.

\bibitem[{Holman} and {Wisdom}, 1993]{HoWi1993}
{Holman}, M.~J. and {Wisdom}, J. (1993).
\newblock {Dynamical stability in the outer solar system and the delivery of
  short period comets}.
\newblock {\em Astron. J.}, 105:1987--1999.

\bibitem[{Hou} et~al., 2014]{HoScLi2014}
{Hou}, X.~Y., {Scheeres}, D.~J., and {Liu}, L. (2014).
\newblock {Saturn Trojans: a dynamical point of view}.
\newblock {\em MNRAS}, 437:1420--1433.

\bibitem[{Izidoro} et~al., 2017]{IzOgRaMo2017}
{Izidoro}, A., {Ogihara}, M., {Raymond}, S.~N., {Morbidelli}, A., {Pierens},
  A., {Bitsch}, B., {Cossou}, C., and {Hersant}, F. (2017).
\newblock {Breaking the chains: hot super-Earth systems from migration and
  disruption of compact resonant chains}.
\newblock {\em \mnras}, 470(2):1750--1770.

\bibitem[{Janson}, 2013]{janson2013}
{Janson}, M. (2013).
\newblock {A Systematic Search for Trojan Planets in the Kepler Data}.
\newblock {\em apj}, 774:156.

\bibitem[{Jenkins} et~al., 2010]{Jenkins2010}
{Jenkins}, J.~M., {Caldwell}, D.~A., {Chandrasekaran}, H., {Twicken}, J.~D.,
  {Bryson}, S.~T., {Quintana}, E.~V., {Clarke}, B.~D., {Li}, J., {Allen}, C.,
  {Tenenbaum}, P., {Wu}, H., {Klaus}, T.~C., {Middour}, C.~K., {Cote}, M.~T.,
  {McCauliff}, S., {Girouard}, F.~R., {Gunter}, J.~P., {Wohler}, B., {Sommers},
  J., {Hall}, J.~R., {Uddin}, A.~K., {Wu}, M.~S., {Bhavsar}, P.~A., {Van
  Cleve}, J., {Pletcher}, D.~L., {Dotson}, J.~A., {Haas}, M.~R., {Gilliland},
  R.~L., {Koch}, D.~G., and {Borucki}, W.~J. (2010).
\newblock {Overview of the Kepler Science Processing Pipeline}.
\newblock {\em \apjl}, 713:L87--L91.

\bibitem[{Jenkins} et~al., 2016]{Jenkins2016}
{Jenkins}, J.~M., {Twicken}, J.~D., {McCauliff}, S., {Campbell}, J.,
  {Sanderfer}, D., {Lung}, D., {Mansouri-Samani}, M., {Girouard}, F.,
  {Tenenbaum}, P., {Klaus}, T., {Smith}, J.~C., {Caldwell}, D.~A., {Chacon},
  A.~D., {Henze}, C., {Heiges}, C., {Latham}, D.~W., {Morgan}, E., {Swade}, D.,
  {Rinehart}, S., and {Vanderspek}, R. (2016).
\newblock {The TESS science processing operations center}.
\newblock In {\em Software and Cyberinfrastructure for Astronomy IV}, volume
  9913 of {\em Society of Photo-Optical Instrumentation Engineers (SPIE)
  Conference Series}, page 99133E.

\bibitem[{Kov{\'a}cs} et~al., 2002]{Kovacs2002}
{Kov{\'a}cs}, G., {Zucker}, S., and {Mazeh}, T. (2002).
\newblock {A box-fitting algorithm in the search for periodic transits}.
\newblock {\em \aap}, 391:369--377.

\bibitem[Lagrange, 1772]{Lagrange1772}
Lagrange (1772).
\newblock {\em \OE uvres compl{\`e}tes}.
\newblock Gouthier-Villars, Paris (1869).

\bibitem[{Laughlin} and {Chambers}, 2002]{LauCha2002}
{Laughlin}, G. and {Chambers}, J.~E. (2002).
\newblock {Extrasolar Trojans: The Viability and Detectability of Planets in
  the 1:1 Resonance}.
\newblock {\em Astron. J.}, 124:592--600.

\bibitem[{Lee} and {Peale}, 2002]{LePe2002}
{Lee}, M.~H. and {Peale}, S.~J. (2002).
\newblock {Dynamics and Origin of the 2:1 Orbital Resonances of the GJ 876
  Planets}.
\newblock {\em ApJ}, 567:596--609.

\bibitem[{Leleu} et~al., 2021]{RIVERS1}
{Leleu}, A., {Chatel}, G., {Udry}, S., {Alibert}, Y., {Delisle}, J.~B., and
  {Mardling}, R. (2021).
\newblock {Alleviating the transit timing variation bias in transit surveys. I.
  RIVERS: Method and detection of a pair of resonant super-Earths around
  Kepler-1705}.
\newblock {\em \aap}, 655:A66.

\bibitem[{Leleu} et~al., 2019a]{LeCoAt2019}
{Leleu}, A., {Coleman}, G., and {Ataiee}, S. (2019a).
\newblock {On the stability of the co-orbital resonance under dissipation:
  Application to the evolution in protoplanetary discs}.
\newblock {\em arXiv e-prints}, page arXiv:1901.07640.

\bibitem[{Leleu} et~al., 2022]{RIVERS2}
{Leleu}, A., {Delisle}, J.~B., {Mardling}, R., {Udry}, S., {Chatel}, G.,
  {Alibert}, Y., and {Eggenberger}, P. (2022).
\newblock {Alleviating the Transit Timing Variations bias in transit surveys.
  II. RIVERS: Twin resonant Earth-sized planets around Kepler-1972 recovered
  from Kepler's false positive}.
\newblock {\em arXiv e-prints}, page arXiv:2201.11459.

\bibitem[{Leleu} et~al., 2019b]{Leleu2019}
{Leleu}, A., {Lillo-Box}, J., {Sestovic}, M., {Robutel}, P., {Correia},
  A.~C.~M., {Hara}, N., {Angerhausen}, D., {Grimm}, S.~L., and {Schneider}, J.
  (2019b).
\newblock {Co-orbital exoplanets from close-period candidates: the TOI-178
  case}.
\newblock {\em \aap}, 624:A46.

\bibitem[{Leleu} et~al., 2015]{LeRoCo2015}
{Leleu}, A., {Robutel}, P., and {Correia}, A.~C.~M. (2015).
\newblock {Detectability of quasi-circular co-orbital planets. Application to
  the radial velocity technique}.
\newblock {\em Astron. Astrophys.}, 581:A128.

\bibitem[{Leleu} et~al., 2018]{LeRoCo2017}
{Leleu}, A., {Robutel}, P., and {Correia}, A.~C.~M. (2018).
\newblock {On the coplanar eccentric non-restricted co-orbital dynamics}.
\newblock {\em Celestial Mechanics and Dynamical Astronomy}, 130(3):24.

\bibitem[{Leleu} et~al., 2017]{LeRoCoLi2017}
{Leleu}, A., {Robutel}, P., {Correia}, A.~C.~M., and {Lillo-Box}, J. (2017).
\newblock {Detection of co-orbital planets by combining transit and
  radial-velocity measurements}.
\newblock {\em Astronomy and Astrophysics}, 599:L7.

\bibitem[Levison et~al., 1997]{LevisonSS97}
Levison, H., Shoemaker, E., and Shoemaker, C. (1997).
\newblock The long-term dynamical stability of {J}upiter's {T}rojan asteroids.
\newblock {\em Nature}, 385:42--44.

\bibitem[{Lillo-Box} et~al., 2018a]{LiBo2017}
{Lillo-Box}, J., {Barrado}, D., {Figueira}, P., {Leleu}, A., {Santos}, N.~C.,
  {Correia}, A.~C.~M., {Robutel}, P., and {Faria}, J.~P. (2018a).
\newblock {The TROY project: Searching for co-orbital bodies to known planets.
  I. Project goals and first results from archival radial velocity}.
\newblock {\em \aap}, 609:A96.

\bibitem[{Lillo-Box} et~al., 2018b]{LiBo2018}
{Lillo-Box}, J., {Leleu}, A., {Parviainen}, H., {Figueira}, P., {Mallonn}, M.,
  {Correia}, A.~C.~M., {Santos}, N.~C., {Robutel}, P., {Lendl}, M., {Boffin},
  H.~M.~J., {Faria}, J.~P., {Barrado}, D., and {Neal}, J. (2018b).
\newblock {The TROY project. II. Multi-technique constraints on exotrojans in
  nine planetary systems}.
\newblock {\em \aap}, 618:A42.

\bibitem[{Lyra} et~al., 2009]{LyJo2009}
{Lyra}, W., {Johansen}, A., {Klahr}, H., and {Piskunov}, N. (2009).
\newblock {Standing on the shoulders of giants. Trojan Earths and vortex
  trapping in low mass self-gravitating protoplanetary disks of gas and
  solids}.
\newblock {\em aap}, 493:1125--1139.

\bibitem[{Madhusudhan} and {Winn}, 2009]{MaWi2009}
{Madhusudhan}, N. and {Winn}, J.~N. (2009).
\newblock {Empirical Constraints on Trojan Companions and Orbital
  Eccentricities in 25 Transiting Exoplanetary Systems}.
\newblock {\em \apj}, 693:784--793.

\bibitem[{Marzari} et~al., 2003]{MaTriSch2003}
{Marzari}, F., {Tricarico}, P., and {Scholl}, H. (2003).
\newblock Stability of jupiter trojans investigated using frequency map
  analysis: the matros project.
\newblock {\em MNRAS}, 345:1091--1100.

\bibitem[{Morais} and {Namouni}, 2016]{MoNa2016}
{Morais}, M.~H.~M. and {Namouni}, F. (2016).
\newblock {A numerical investigation of coorbital stability and libration in
  three dimensions}.
\newblock {\em Celestial Mechanics and Dynamical Astronomy}, 125(1):91--106.

\bibitem[{Namouni}, 1999]{Namouni1999}
{Namouni}, F. (1999).
\newblock {Secular Interactions of Coorbiting Objects}.
\newblock {\em Icarus}, 137:293--314.

\bibitem[{Nauenberg}, 2002]{Nauenberg2002}
{Nauenberg}, M. (2002).
\newblock {Stability and Eccentricity for Two Planets in a 1:1 Resonance, and
  Their Possible Occurrence in Extrasolar Planetary Systems}.
\newblock {\em Astron. J.}, 124:2332--2338.

\bibitem[Nesvorny and Dones, 2002]{NeDo02}
Nesvorny, D. and Dones, L. (2002).
\newblock How long-live are the hypothetical {T}rojan populations of {S}aturn,
  {U}ranus, and {N}eptune?
\newblock {\em Icarus}, 160:271--288.

\bibitem[{Nesvorn{\'y}} et~al., 2013]{Nesvorny2013}
{Nesvorn{\'y}}, D., {Kipping}, D., {Terrell}, D., {Hartman}, J., {Bakos},
  G.~{\'A}., and {Buchhave}, L.~A. (2013).
\newblock {KOI-142, The King of Transit Variations, is a Pair of Planets near
  the 2:1 Resonance}.
\newblock {\em \apj}, 777(1):3.

\bibitem[{Nesvorn{\'y}} and {Vokrouhlick{\'y}}, 2014]{NeVo2014}
{Nesvorn{\'y}}, D. and {Vokrouhlick{\'y}}, D. (2014).
\newblock {The Effect of Conjunctions on the Transit Timing Variations of
  Exoplanets}.
\newblock {\em \apj}, 790(1):58.

\bibitem[{Nesvorn{\'y}} and {Vokrouhlick{\'y}}, 2016]{NeVo2016}
{Nesvorn{\'y}}, D. and {Vokrouhlick{\'y}}, D. (2016).
\newblock {Dynamics and Transit Variations of Resonant Exoplanets}.
\newblock {\em \apj}, 823(2):72.

\bibitem[{Nesvorny} et~al., 2013]{NeVoMo2013}
{Nesvorny}, D., {Vokrouhlicky}, D., and {Morbidelli}, A. (2013).
\newblock {Capture of Trojans by Jumping Jupiter}.
\newblock {\em APJ}, 768:45.

\bibitem[{Papaloizou} and {Terquem}, 2010]{PaTe2010}
{Papaloizou}, J. C.~B. and {Terquem}, C. (2010).
\newblock {On the dynamics of multiple systems of hot super-Earths and
  Neptunes: tidal circularization, resonance and the HD 40307 system}.
\newblock {\em \mnras}, 405(1):573--592.

\bibitem[{Penzlin} et~al., 2019]{PeAtKl2019}
{Penzlin}, A. B.~T., {Ataiee}, S., and {Kley}, W. (2019).
\newblock {1:1 orbital resonance of circumbinary planets}.
\newblock {\em \aap}, 630:L1.

\bibitem[{Petit} et~al., 2018]{PeLaBo2018}
{Petit}, A.~C., {Laskar}, J., and {Bou{\'e}}, G. (2018).
\newblock {Hill stability in the AMD framework}.
\newblock {\em \aap}, 617:A93.

\bibitem[{Pierens} and {Raymond}, 2014]{PiRa2014}
{Pierens}, A. and {Raymond}, S.~N. (2014).
\newblock {Disruption of co-orbital (1:1) planetary resonances during
  gas-driven orbital migration}.
\newblock {\em mnras}, 442:2296--2303.

\bibitem[{Pousse} et~al., 2017]{PoRoVi2017}
{Pousse}, A., {Robutel}, P., and {Vienne}, A. (2017).
\newblock {On the co-orbital motion in the planar restricted three-body
  problem: the quasi-satellite motion revisited}.
\newblock {\em Celestial Mechanics and Dynamical Astronomy}.

\bibitem[{Pu} and {Wu}, 2015]{PuWu2015}
{Pu}, B. and {Wu}, Y. (2015).
\newblock {Spacing of Kepler Planets: Sculpting by Dynamical Instability}.
\newblock {\em \apj}, 807:44.

\bibitem[{Qi} and {de Ruiter}, 2020]{QiRu2020}
{Qi}, Y. and {de Ruiter}, A. (2020).
\newblock {Phase Structure of Co-orbital Motion with Jupiter}.
\newblock {\em \mnras}.

\bibitem[{Roberts}, 2002]{Robe2002}
{Roberts}, G. (2002).
\newblock Linear stability of the elliptic {L}agrangian triangle solutions in
  thethree-body problem.
\newblock {\em JDIFE}, 182:191--218.

\bibitem[{Robutel} and {Gabern}, 2006]{RoGa2006}
{Robutel}, P. and {Gabern}, F. (2006).
\newblock The resonant structure of {J}upiter's {T}rojan asteroids {I}:
  Long-term stability and diffusion.
\newblock {\em MNRAS}, 372:1463--1482.

\bibitem[Rodr{\'\i}guez et~al., 2013]{RoGiMi2013}
Rodr{\'\i}guez, A., {Giuppone}, C.~A., and {Michtchenko}, T.~A. (2013).
\newblock Tidal evolution of close-in exoplanets in co-orbital configurations.
\newblock {\em Celest. Mech. Dyn. Astron.}, 117:59--74.

\bibitem[{Sidorenko} et~al., 2014]{SiArNeZe2013}
{Sidorenko}, V.~V., {Neishtadt}, A.~I., {Artemyev}, A.~V., and {Zelenyi}, L.~M.
  (2014).
\newblock {Quasi-satellite orbits in the general context of dynamics in the 1:1
  mean motion resonance: perturbative treatment}.
\newblock {\em Celestial Mechanics and Dynamical Astronomy}, 120(2):131--162.

\bibitem[{Terquem} and {Papaloizou}, 2007]{TePa2007}
{Terquem}, C. and {Papaloizou}, J. C.~B. (2007).
\newblock {Migration and the Formation of Systems of Hot Super-Earths and
  Neptunes}.
\newblock {\em \apj}, 654(2):1110--1120.

\bibitem[{Thomas} et~al., 2013]{ThomasBurnsHedmanHelfensteinMorrison2013}
{Thomas}, P.~C., {Burns}, J.~A., Hedman, M., {Helfenstein}, P., Morrison, S.,
  Tiscareno, and {Veverka}, J. (2013).
\newblock The inner small satellites of saturn: A variety of worlds.
\newblock {\em Icarus}, 226(999--1019).

\bibitem[{Vokrouhlick{\'y}} and {Nesvorn{\'y}}, 2014]{VoNe2014}
{Vokrouhlick{\'y}}, D. and {Nesvorn{\'y}}, D. (2014).
\newblock {Transit Timing Variations for Planets Co-orbiting in the Horseshoe
  Regime}.
\newblock {\em ApJ}, 791:6.

\bibitem[{Wisdom}, 1980]{Wi1980}
{Wisdom}, J. (1980).
\newblock {The resonance overlap criterion and the onset of stochastic behavior
  in the restricted three-body problem}.
\newblock {\em Astrophysical Journal}, 85:1122--1133.

\bibitem[{Wolf}, 1906]{Wolf1906}
{Wolf}, M. (1906).
\newblock {Photographische Aufnahmen von kleinen Planeten}.
\newblock {\em Astronomische Nachrichten}, 170:353.

\end{thebibliography}

\end{document}